\documentclass[letterpaper, 10 pt, conference]{ieeeconf} % Comment this line out if you need a4paper

\IEEEoverridecommandlockouts % This command is only needed if
\usepackage{graphics} % for pdf, bitmapped graphics files
\usepackage{epsfig} % for postscript graphics files
\usepackage{mathptmx} % assumes new font selection scheme installed
\usepackage{times} % assumes new font selection scheme installed
\usepackage{amsmath} % assumes amsmath package installed
\usepackage{amssymb} % assumes amsmath package installed
\usepackage{subfig} % for subfloat command

\title{Modeling and control of a low-cost multirotor hybrid aerial underwater vehicle
}
\author{RenKai Wang, ZhiGang Shang}

\begin{document}

\maketitle
\pagestyle{empty}
\thispagestyle{empty}

\begin{abstract}
This paper presents a comprehensive modeling and control framework for a low-cost multirotor hybrid aerial-aquatic vehicle (MHAUV) capable of seamless air-water transitions. A hybrid dynamics model is proposed to account for the distinct hydrodynamic and aerodynamic forces across three operational zones: aerial, aquatic, and transitional hybrid regions. The model incorporates variable buoyancy, added mass effects, and fluid resistance, with thrust characteristics of submerged propellers analyzed through computational fluid dynamics (CFD) simulations. A hierarchical control strategy is developed, combining twisting sliding mode control (TWSMC) for robust attitude stabilization during medium transitions with cascade PID controllers for precise motion tracking in homogeneous media. Experimental validation using a modified FPV quadrotor prototype demonstrates the effectiveness of the approach, achieving steady-state height errors below 0.1 m and attitude fluctuations under 5° during repeated water-crossing maneuvers. The results highlight the system's adaptability to fluid medium variations while maintaining cost-effectiveness and operational simplicity.

\end{abstract}

\section{Introduction}
The exploration and monitoring of complex environments often require robotic platforms capable of operating in multiple domains.\cite{zeng2022review,yao2023review} Hybrid aerial-underwater vehicles (HAUVs), also known as aerial-aquatic vehicles, have emerged as a promising solution that offers the potential to bridge the gap between aerial surveillance and underwater inspection.\cite{xiang2015hybrid} Multirotor configurations, specifically quadrotors, are particularly attractive for HAUV development due to their vertical take-off and landing (VTOL) capabilities, high maneuverability in air, mechanical simplicity, and relatively low cost compared to fixed-wing or more complex biomimetic designs.\cite{bi2022nezha} These characteristics make them suitable for a wide range of applications, including search and rescue operations, environmental monitoring of water bodies, bridge and dam inspection, and maintenance of underwater structures.
Despite their potential, the development of effective multirotor HAUVs (MHAUVs) faces significant challenges, primarily centered on the air-water transition phase. The abrupt and drastic change in fluid properties, approximately 800 times the difference in density and 50 times the difference in viscosity between water and air, imposes severe demands on vehicle design, modeling, and control. Key difficulties include:
(i) maintaining stable flight and buoyancy control during entry and exit;\cite{alzubi2018loon,bi2022dynamics}
(ii) ensuring adequate thrust generation from propellers designed primarily for air when operating submerged or partially submerged;\cite{alzubi2015evaluation,horn2019study}
(iii) accurately modeling the complex, time-varying hydrodynamic and aerodynamic forces, including added mass effects, slamming forces, and varying buoyancy during the transition;\cite{drews2014hybrid,mercado2018modeling}
(iv) developing robust control strategies capable of handling these rapid dynamic changes and external disturbances like waves or wind gusts.\cite{neto2015attitude,farinha2021challenges}

Previous research has explored various approaches to address these challenges. Some designs incorporate complex transition mechanisms, such as folding wings or dedicated underwater thrusters, which can increase weight, complexity, and cost.\cite{bai2023nezha,bi2024design} Others have focused on fixed-wing HAUVs, which offer efficiency but lack hovering and low-speed maneuverability of multirotors.\cite{jin2025nezha,li2022aerial} For multirotor HAUVs, modeling efforts often simplify the transition dynamics or assume quasistatic conditions.\cite{chen2019system,horn2020novel} Although various control techniques, including PID,\cite{zongcheng2022ga} LQR, adaptive control,\cite{lu2020adaptive} and sliding mode control (SMC),\cite{chen2020attitude} have been applied, achieving robust stability and performance \textit{during} the highly nonlinear transition phase, especially for low-cost systems with potentially noisy sensors and unmodeled dynamics, remains an active area of research. Furthermore, accurately characterizing propeller performance across the air-water interface, which is crucial for both modeling and control, often requires detailed analysis beyond standard propeller theory.\cite{semenovnddevelopment,bai2025review}

This paper addresses the critical need for accurate modeling and robust control of low-cost multirotor HAUVs capable of seamless air-water transitions. Our primary objective is to develop and validate a comprehensive framework that enables reliable cross-domain operation using a cost-effective platform derived from readily available components. The main contributions of this work are threefold.

A hybrid dynamics model is proposed that explicitly considers three distinct operational zones: aerial, aquatic, and a transitional hybrid region near the water surface. This model incorporates variable buoyancy, added mass effects, and fluid resistance that change depending on the vehicle's immersion depth. The thrust characteristics of propellers operating across the air-water interface are systematically investigated using computational fluid dynamics (CFD) simulations. This analysis provides a data-driven relationship between propeller immersion depth and the resulting thrust coefficient, enhancing the model fidelity. A hierarchical control strategy is developed, employing twisting sliding mode control (TWSMC) specifically for robust attitude stabilization during the challenging medium transition phase, complemented by cascade PID controllers for precise motion tracking in homogeneous aerial and aquatic zones. This hybrid approach aims to balance robustness against transient disturbances with smooth steady-state performance. The effectiveness of the proposed modeling and control framework is validated through both simulations and experiments using a low-cost prototype adapted from a commercial FPV quadrotor, demonstrating its practical feasibility and performance in repeated water-crossing maneuvers.

\section{Dynamic Model}\label{sec:Dynamic Model}

\subsection{Fluid Zone Definition}
\begin{figure}[htpb!]
    \centering
    \includegraphics[width=0.5\linewidth]{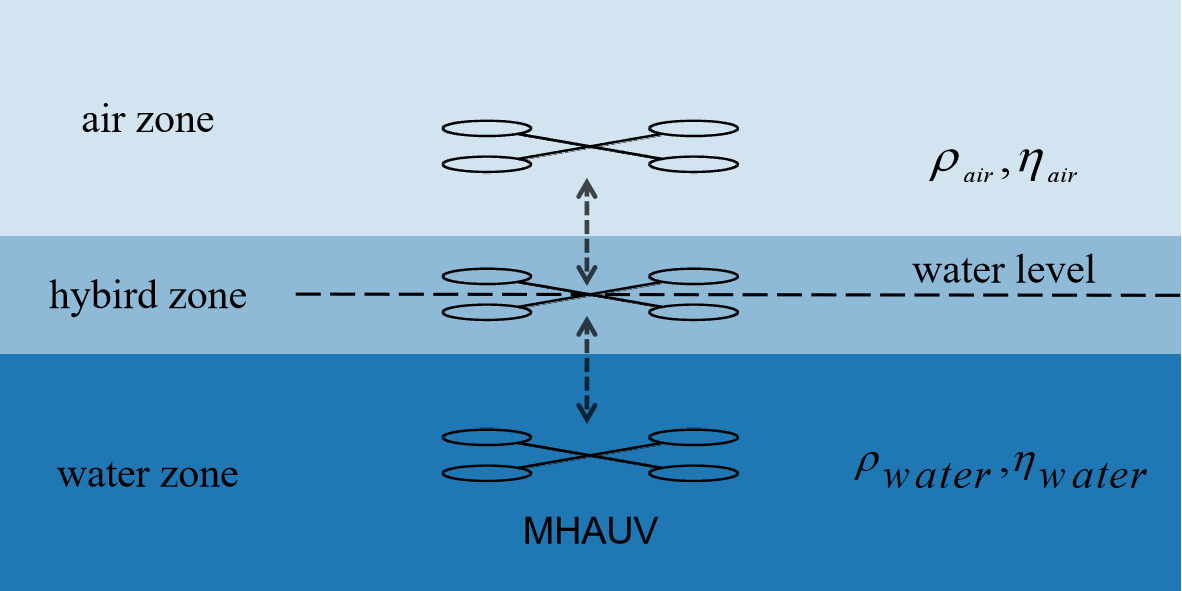}
    \caption{Zone define}
    \label{zone difine}
\end{figure}
The characteristics of the fluid region surrounding the MHAUV undergo significant changes as it crosses the water surface, resulting in notable alterations in its forces.
Due to the vehicle's finite vertical dimension (height $H$), this transition is not instantaneous, but occurs over a range of depths. To capture the distinct dynamics, we define three operational zones based on the vertical position z of the vehicle's center of gravity relative to the water surface (defined as z = 0), as illustrated in Figure \ref{zone difine}:

$\bullet$ Air Zone ($z \geq H/2$): The vehicle is fully in the air. Aerodynamic forces dominate, and hydrodynamic effects are negligible.

$\bullet$ Water Zone ($z \leq -H/2$): The vehicle is fully submerged in water. Hydrodynamic forces (added mass, buoyancy, drag) are dominant.

$\bullet$ Hybrid Zone ($-H/2 < z < H/2$): The vehicle is partially submerged, crossing the air-water interface. Both aerodynamic and hydrodynamic forces are significant and vary rapidly with immersion depth.

\subsection{Aerial Dynamics}
\begin{figure}[htpb!]
    \centering
    \includegraphics[width=0.5\linewidth]{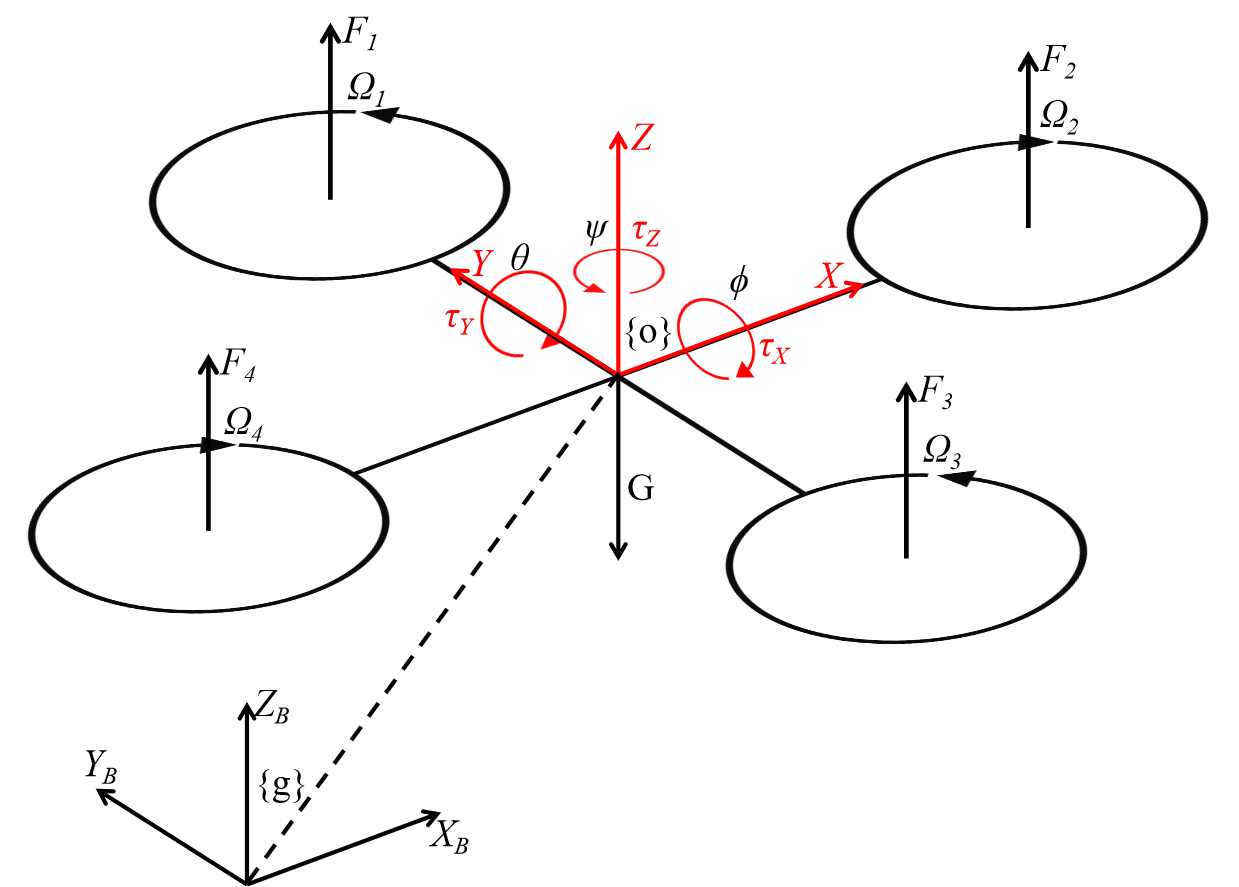}
    \caption{Coordinate define}
    \label{coordinate}
\end{figure}

We consider a rigid quadrotor MHAUV as shown in \ref{coordinate}. The adjacent arms are of equal length = $l$ and are positioned perpendicularly to each other, with the motor and propeller at the end of the arms and the control unit in the center.
The propellers are numbered 1 to 4, and the adjacent propellers rotate in opposite directions at a speed of $\Omega_i$ to counteract a part of the torque.
The thrust $F_i$ generated by the propeller i is aligned parallel to the Z axis within the fuselage coordinate system and is non-negative. Furthermore, the moment $M_i$ resulting from resistance acts in opposition to the direction of rotation.

This paper examines the MHAUV operates at low speeds and the air density $\rho_{air}$ is significantly low. Consequently, it is assumed that the effects of air resistance $R_{air}$, air buoyancy $B_{air}$, and additional mass $m_{air}$ are nearly zero. The aerial dynamics model of the MHAUV is as follows:
\begin{gather}\label{eq1}
    \begin{bmatrix}
    \dot{x} \\
    \dot{y} \\
    \dot{z}
    \end{bmatrix}=
    \begin{bmatrix}
    c_\theta c_\psi & s_\phi s_\theta c_\psi-c_\phi s_\psi & c_\phi s_\theta c_\psi+s_\phi s_\psi \\
    c_\theta s_\psi & s_\phi s_\theta s_\psi+c_\phi c_\psi & c_\phi s_\theta s_\psi-s_\phi c_\psi \\
    -s_\theta & s_\phi c_\theta & c_\phi c_\theta
    \end{bmatrix}
    \begin{bmatrix}
    u \\
    v \\
    w
    \end{bmatrix},
\end{gather}
\begin{gather}\label{eq:dyn_trans_air}
    \begin{bmatrix}
    \dot{u} \\
    \dot{v} \\
    \dot{w}
    \end{bmatrix}=
    \begin{bmatrix}
    0 & -R & Q \\
    R & 0 & -P \\
    -Q & P & 0
    \end{bmatrix}
    \begin{bmatrix}
    u \\
    v \\
    w
    \end{bmatrix}+
    \begin{bmatrix}
    g\sin\theta \\
    -g\sin\phi\cos\theta \\
    -g\cos\phi\cos\theta+\frac{T_z}{m}
    \end{bmatrix},
\end{gather}
\begin{gather}\label{eq:dyn_rot_air}
    \begin{split}
        \begin{bmatrix}
        \dot{P} \\
        \dot{Q} \\
        \dot{R}
        \end{bmatrix}=
        \begin{bmatrix}
        0 & \frac{I_{yy}-I_{zz}}{I_{xx}}R & \frac{I_{yy}-I_{zz}}{I_{xx}}Q \\
        \frac{I_{xz}-I_{xz}}{I_{yy}}R & 0 & \frac{I_{xz}-I_{xz}}{I_{yy}}P \\
        \frac{I_{xz}-I_{yy}}{I_{zz}}Q & \frac{I_{xz}-I_{yy}}{I_{zz}}P & 0
        \end{bmatrix}
        \begin{bmatrix}
        P \\
        Q \\
        R
        \end{bmatrix}\\
        +
        \begin{bmatrix}
        \frac{1}{I_{xx}} & 0 & 0 \\
        0 & \frac{1}{I_{yy}} & 0 \\
        0 & 0 & \frac{1}{I_{zz}}
        \end{bmatrix}
        \begin{bmatrix}
        \tau_x \\
        \tau_y \\
        \tau_z
        \end{bmatrix}+
        \begin{bmatrix}
        -\frac{I_{xzx}}{I_{xx}}QW_G \\
        -\frac{I_{xzx}}{I_{yy}}PW_G \\
        0
        \end{bmatrix},
    \end{split}
\end{gather}
\begin{gather}\label{eq4}
    \begin{bmatrix}
    \dot{\phi} \\
    \dot{\theta} \\
    \dot{\psi}
    \end{bmatrix}=
    \begin{bmatrix}
    1 & s_\phi t_\theta & c_\phi t_\theta \\
    0 & c_\phi & -s_\phi \\
    0 & \frac{s_\phi}{c_\theta} & \frac{c_\phi}{c_\theta}
    \end{bmatrix}
    \begin{bmatrix}
        p\\
        q\\
        r
    \end{bmatrix},
\end{gather}
\begin{gather}
    T_z=T_1+T_2+T_3+T_4,\\
    \tau_x=T_1l-T_3l,\\
    \tau_y=-T_2l+T_4l\,\\
    \tau_z=-M_1+M_2-M_3+M_4,
\end{gather}

In the above equations, the position coordinates of the MHAUV in the inertial frame are described by $x, y, z$.
Meanwhile, the velocity components in the body frame are defined as $u, v, w$.
The orientation of the MHAUV is described by the roll ($\phi$), pitch ($\theta$), and yaw ($\psi$) angles, while the angular velocity components around the body frame axes are given by $P, Q, R$.
The total thrust force acting in the $z$-direction of the body frame is $T_z$, and the moments (torques) around the body frame axes are denoted by $\tau_x, \tau_y, \tau_z$.
The mass of the MHAUV is denoted by $m$, and the moments of inertia around the body frame axes are $I_{xx}, I_{yy}, I_{zz}$. The acceleration due to gravity is represented by $g$.
Specifically, $I_{zzm}$ represents the moment of inertia related to the thrusters, while $W_G$ denotes the angular velocity of the thrusters.
Additionally, the aforementioned model is grounded in a generalized framework designed for the modeling of quadrotor aerial vehicles.
% (Tahirovic et al., 2024)

\subsection{Aquatic Dynamics}
When fully submerged, the MHAUV experiences significant hydrodynamic forces that are negligible in air. The primary effects are added mass, buoyancy, and hydrodynamic drag/damping.\cite{drews2014hybrid}
\begin{gather}
    m=m_{v}+m_{a},\\
    g=g_0-g_B=g_0-B/m_v,
\end{gather}

where $m_v$ and $m_a$ denote the mass of the vehicle and the added mass respectively, while $g_0$ and $g_B$ represent the acceleration due to gravity and buyacy $B$.

Due to the relatively low speed of the MHAUV, drag force $F_d$ and the moment $M_d$ are estimated as follows.
\begin{gather}
    \begin{bmatrix}
        F_d \\
        M_d
    \end{bmatrix}=\frac{1}{2}C_dA_d\rho\nu\circ \nu,\\
    F_d=[F_{du},F_{dv},F_{dw}]^T,\\
    M_d=[M_{d\phi},M_{d\theta},M_{d\psi}]^T,\\
    A_d=diag[A_{du},A_{dv},A_{dw},A_{d\phi},A_{d\theta},A_{d\psi}],\\
    \nu=[u,v,w,\phi,\theta,\psi]^T,
\end{gather}

where the vectors $F_d$ and $M_d$ represent the three-dimensional values of water resistance.  The drag coefficient $C_d$ is influenced by the shape and surface roughness of the MHAUV. The diagonal elements of the projected area matrix $A_d$ are respectively utilized to describe the projected area of MHAUV in various directions, with direction vectors defined as 6-DOF translations and rotations along the axes of the body frame.

Incorporating these terms (using the modified mass m and effective gravity g) modifies the translational and rotational dynamic equations (\ref{eq:dyn_trans_air}) and (\ref{eq:dyn_rot_air}):
\begin{gather}
    \begin{split}
    \begin{bmatrix}
    \dot{u} \\
    \dot{v} \\
    \dot{w}
    \end{bmatrix}
    &=
    \begin{bmatrix}
    0 & -R & Q \\
    R & 0 & -P \\
    -Q & P & 0
    \end{bmatrix}
    \begin{bmatrix}
    u \\
    v \\
    w
    \end{bmatrix}\\
    &+
    \begin{bmatrix}
    g\sin\theta \\
    -g\sin\phi\cos\theta \\
    -g\cos\phi\cos\theta+\frac{T_z}{m}
    \end{bmatrix}-
    \begin{bmatrix}
    \frac{F_{du}}{m} \\
    \frac{F_{dv}}{m} \\
    \frac{F_{dv}}{m}
    \end{bmatrix},
    \end{split}
\end{gather}
\begin{equation}
        \begin{split}
\begin{bmatrix}
\dot{P} \\
\dot{Q} \\
\dot{R}
\end{bmatrix}
=
\begin{bmatrix}
0 & \frac{I_{yy} - I_{zz}}{I_{xx}} R & \frac{I_{yy} - I_{zz}}{I_{xx}} Q \\
\frac{I_{zz} - I_{xx}}{I_{yy}} R & 0 & \frac{I_{zz} - I_{xx}}{I_{yy}} P \\
\frac{I_{xx} - I_{yy}}{I_{zz}} Q & \frac{I_{xx} - I_{yy}}{I_{zz}} P & 0
\end{bmatrix}
\begin{bmatrix}
P \\
Q \\
R
\end{bmatrix}
+\\
\begin{bmatrix}
\frac{1}{I_{xx}} & 0 & 0 \\
0 & \frac{1}{I_{yy}} & 0 \\
0 & 0 & \frac{1}{I_{zz}}
\end{bmatrix}
\begin{bmatrix}
\tau_x \\
\tau_y \\
\tau_z
\end{bmatrix}
+
\begin{bmatrix}
-\frac{I_{xzx}}{I_{xx}} Q W_G \\
-\frac{I_{xzx}}{I_{yy}} P W_G \\
0
\end{bmatrix}
-
\begin{bmatrix}
\frac{M_\phi}{I_{xx}} \\
\frac{M_\theta}{I_{yy}} \\
\frac{M_\psi}{I_{zz}}
\end{bmatrix}
        \end{split}
\end{equation}

\subsection{Hybrid Dynamics}
The hybrid zone (-H/2 < z < H/2) represents the challenging transition phase where the vehicle is partially submerged. The hydrodynamic forces are no longer constant but vary with the degree of immersion.
To model this, we introduce depth-dependent weighting coefficients $C_m$, $C_B$ and $C_A$ that scale the fully submerged added mass $m_a$, buoyancy $B$ and drag areas $A_d$, respectively:
\begin{gather}
    m_a=C_m m_{a0},\\
    B=C_B\rho_{w}g_0V_{0},\\
    A_d=C_A A_{d0},
\end{gather}

These coefficients represent the ratio of the effect at a given depth z compared to the fully submerged case. A crucial simplifying assumption is made: The vehicle maintains a near-vertical orientation (small roll $\phi$ and pitch $\theta$ angles) during the transition.
This allows the coefficients to be approximated as functions solely of the vertical position z of the center of gravity. A simple linear weighting function is adopted:
\begin{gather}
    C_m=C_B=C_A=
    \begin{cases}
        0,\quad z\geq\frac{1}{2}H\\
        0.5 - \frac{z}{H},\quad -\frac{1}{2}H<z<\frac{1}{2}H\\
        1,\quad z\leq-\frac{1}{2}H\\
    \end{cases}
\end{gather}

It is important to recognize the limitations of this linear weighting approach. The actual physics of air-water transition are highly complex and nonlinear, involving effects like surface tension, slamming forces upon impact, ventilation, splashing, and the intricate near-water effect influencing propeller performance.\cite{bai2025review}

Other research efforts have also employed linearization techniques for transition dynamics, often treating the resulting linearization errors, along with phenomena like ground effect, as disturbances to be handled by the controller.\cite{mercado2018modeling} The justification for this simplification in our work stems from the low-cost nature of the platform, where developing a highly complex, experimentally validated nonlinear transition model is prohibitive.

Instead, the focus shifts towards designing a robust control system (Section 3) capable of compensating for the inevitable inaccuracies introduced by this simplified hybrid model. The validity of assuming near-vertical orientation during transition is contingent on the attitude controller's ability to maintain stability, which is verified experimentally in Section 4. If the controller were unable to keep attitude fluctuations small, the accuracy of this hybrid model component would degrade significantly.

\subsection{Propeller Dynamics}
The thrust generated by a propeller is fundamentally affected by the density of the medium it operates in.\cite{alzubi2015evaluation}
In typical scenarios, the thrust produced by the propeller $T$ is described by the following equation:
\begin{equation}\label{eq:thrust_prop}
    T = C_T\Omega^2D^4,
\end{equation}

where the thrust coefficient $C_T$ is determined by the pitch angle $\theta_p$, while also being influenced by the surrounding environmental medium, such as water and air.
In this paper, we assume that this effect of the medium on the propeller thrust coefficient can be simplified as a function of the height difference $h$ between the water level and the center of the propeller.
In addition, the power factors associated with the rotational speed $\Omega$ and diameter $D$ are generally assumed constant values of 2 and 4 respectively.

\begin{figure}
    \centering
    \includegraphics[width=0.5\linewidth]{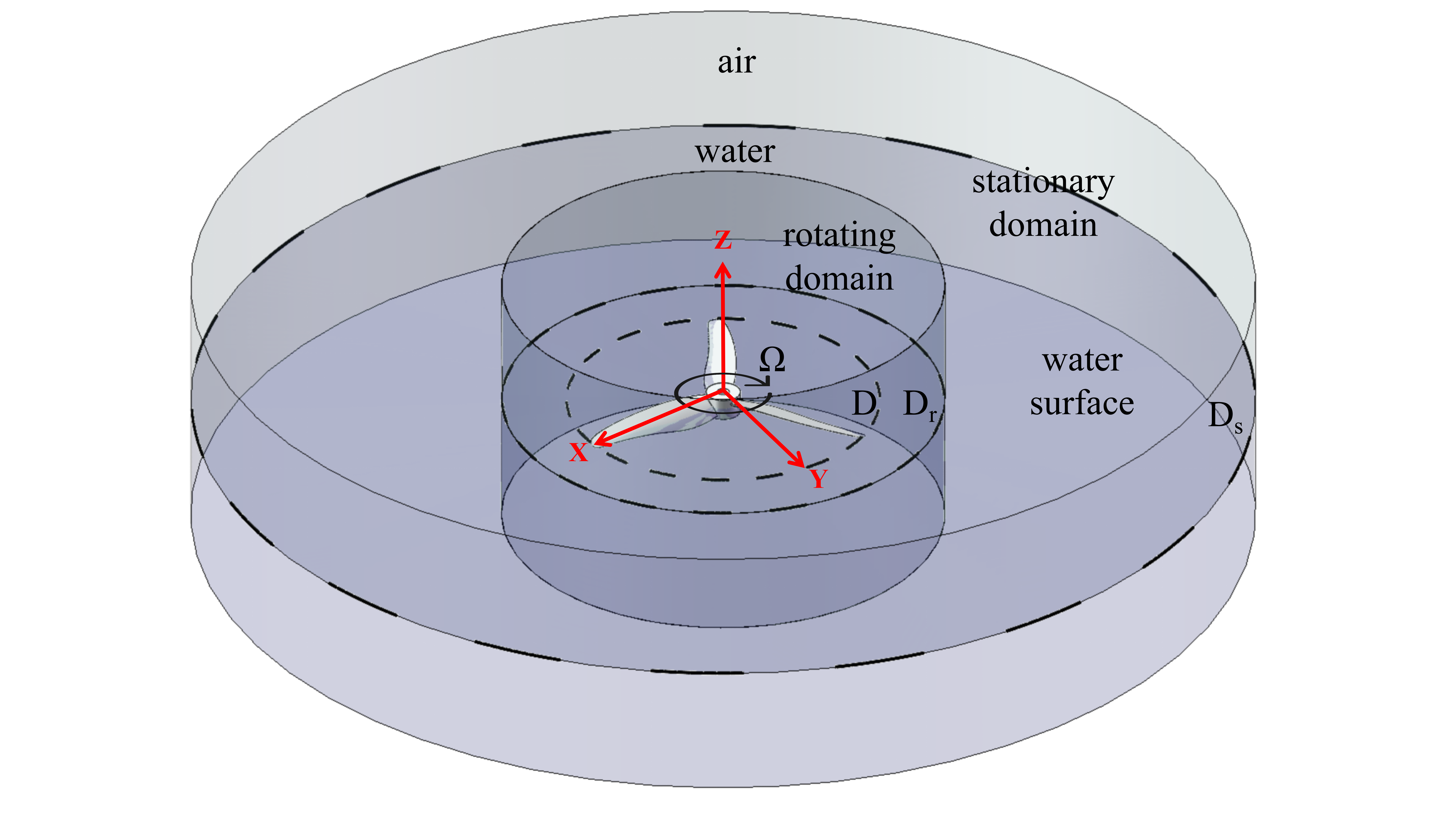}
    \caption{CFD computing region}
    \label{cfd define}
\end{figure}

In this paper, computer fluid dynamics(CFD) simulations are used to calculate the relationship between $C_T$ and $h$.
The computing domain is illustrated in \ref{cfd define}, where the propeller of diameter $D$ rotates around the Z axis at its center. The computing domain is separated by the rotating domain of diameter $D_r$ and the stationary domain of diameter $D_s$.

To facilitate calculations, we assume that the center of the propeller is positioned at the origin of the coordinate system, with the thrust direction of the propeller aligned along the positive z-axis. Consequently, the plane representing the water surface is defined as \( z = h \).

The CFD simulations utilize Fluent as the basis, employing the shear stress transport (sst) model along with the Euler-Euler approach for conducting transient simulations.
In order to reduce the consumption of computing resources and time, we use the steady-state model at the beginning of the simulation, and then switch to the transient model when it runs smoothly.

The principal parameters and variables involved are listed in \ref{cfd table}, where the combination of $\Omega$, $D$ and $h$ constitutes 500 situations.

\begin{table}[!t]
\caption{Parameters used in CFD}
    \centering
    \begin{tabular}{|c||c|}
    \hline
    $\Omega(rpm)$ & 100, 250, 500, 1000, 1500, 2000\\
    \hline
    D(inch) & 3.5, 5, 7, 10\\
    \hline
    h(mm) & 0, $\pm20$, $\pm40$, $\pm60$, $...$, $\pm200$\\
    \hline
    $D_r/D$& 2.5\\
    \hline
    $D_s/D$& 25\\
    \hline
    \end{tabular}
    \label{cfd table}
\end{table}

For situations of D=3.5, we use the GEMFAN3520 three-blade propeller, which is identical to the prototype model. For other conditions, we utilize this model and scale it up to mitigate the effects of propeller profile and pitch angle.

We assume that the thrust $T$ is equal to the axial component of the integral of the propeller surface force. In transient simulations, this force exhibits periodic variations; thus, we take its average value over one period as the thrust value.

\begin{figure}
    \centering
    \includegraphics[width=0.5\linewidth]{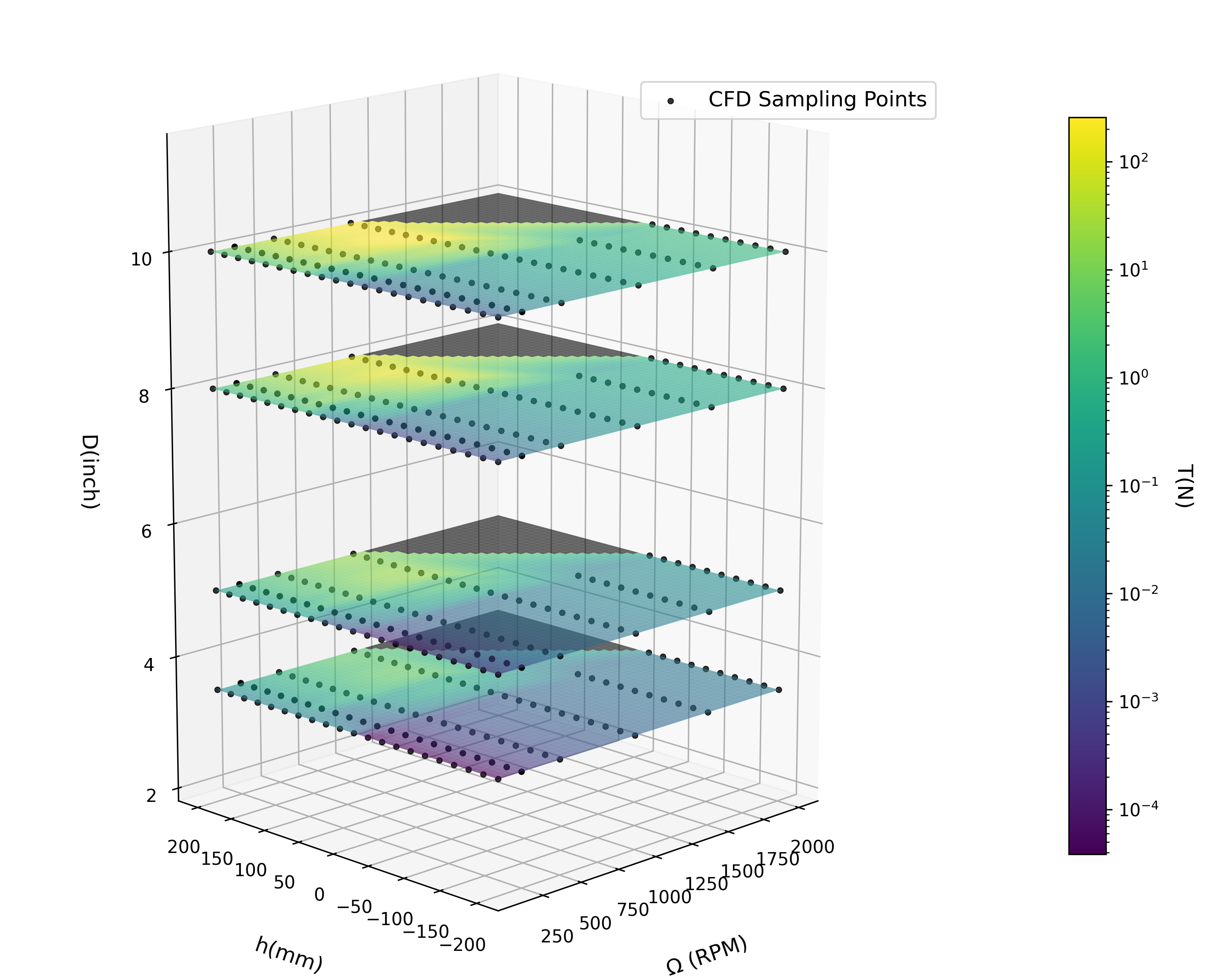}
    \caption{Scatter diagram of working conditions}
    \label{cfd result1}
\end{figure}

\begin{figure}
    \centering
    \includegraphics[width=0.5\linewidth]{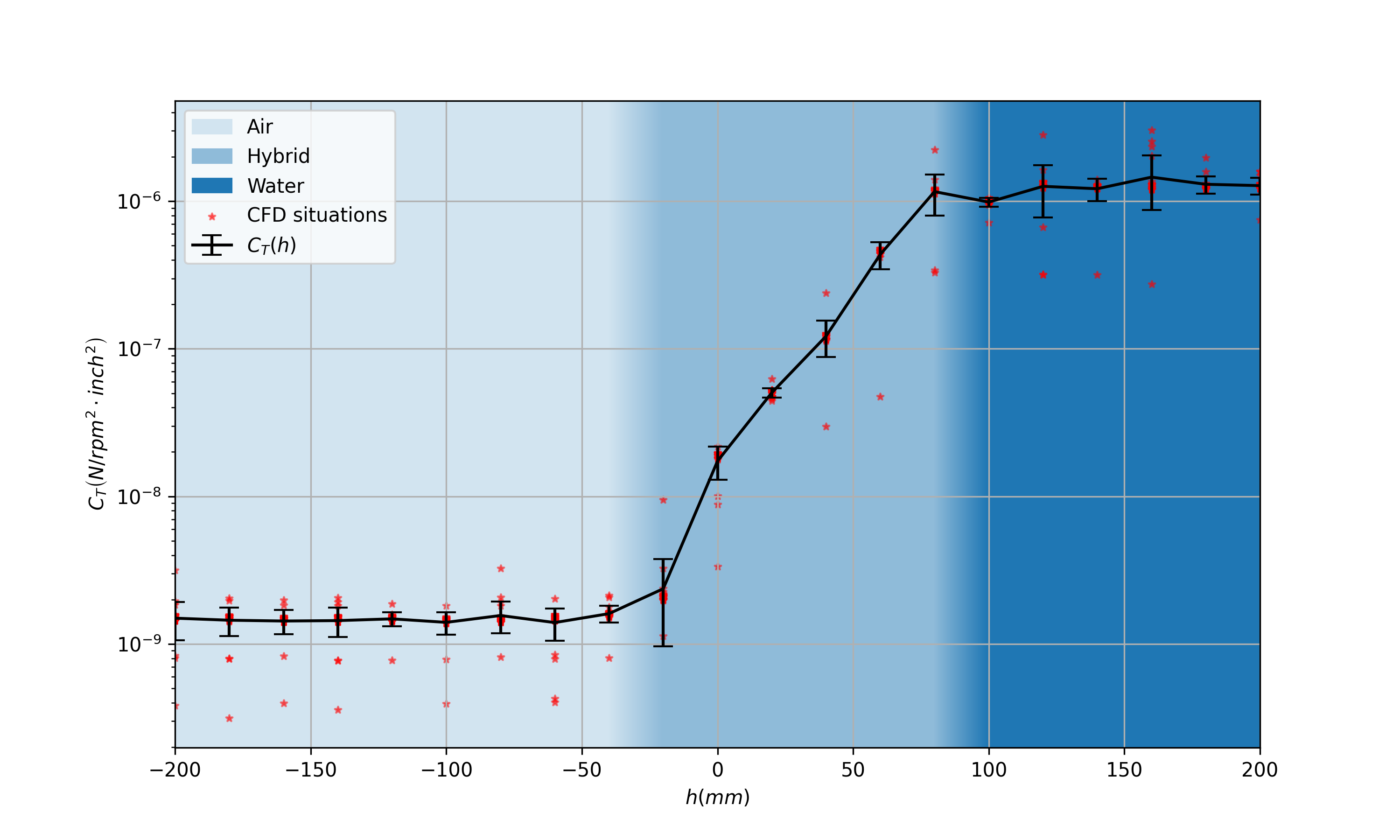}
    \caption{$C_T-h$ curve in air, hybrid and water zone}
    \label{C_T result}
\end{figure}

% \begin{figure}
%     \centering
%     \includegraphics[width=0.3\linewidth]{volume.png}
%     \caption{Top view of water-air phase(red: air, blue: water) distribution.}
%     \label{volume}
% \end{figure}

The distribution of h-Omega-D under all CFD conditions is illustrated in \ref{cfd result1}, where the thermal map of the D plane represents the thrust distribution of a propeller with a specific diameter. The CFD simulations confirmed that thrust $T$ increases significantly with immersion depth $h$, rotational speed $\Omega$, and propeller diameter $D$ (Figure \ref{cfd result1}).

Using the simulation results and the chosen thrust equation form Eq. \ref{eq:thrust_prop}, the thrust coefficient $C_T(h)$ was calculated as a function of immersion $h$. The resulting $C_T$ curve is shown in Figure \ref{C_T result}.

Among them, as shown in the air area, when $h<-50mm$, $C_T$ is stable at about 1.5*10-9; , as shown in the water zone, When $h>100mm$, $C_T$ size stabilizes to about 1.3*10-6. In addition, we note that $logC_T-h$ are approximately linear in the hybrid zone. Therefore, $C_T(h)$ is simplified to the following equation:

\begin{equation}
    C_T=\begin{cases}
        C_{Ta}=1.5\times10-9,\quad h<-50mm,\\
        C_{Tw}=1.3\times10-6,\quad h>100mm,\\
        e^{\alpha \cdot log_{C_{Ta}}+(1-\alpha)\cdot log_{C_{Tw}}},\quad else,
    \end{cases}
\end{equation}

where the constants $C_{Ta}$ and $C_{Tw}$ refers to the thrust coefficient in air and water, respectively. The weighting factor $\alpha$ is calculated as $\alpha=(100mm-h(mm))/150mm$.This data-driven model for $C_T(h)$ provides a crucial component for accurately predicting propeller thrust during the transition, directly addressing one of the key challenges outlined in the Introduction.

\section{Control Strategy}\label{sec:Control Strategy}
\begin{figure}
    \centering
    \includegraphics[width=0.5\linewidth]{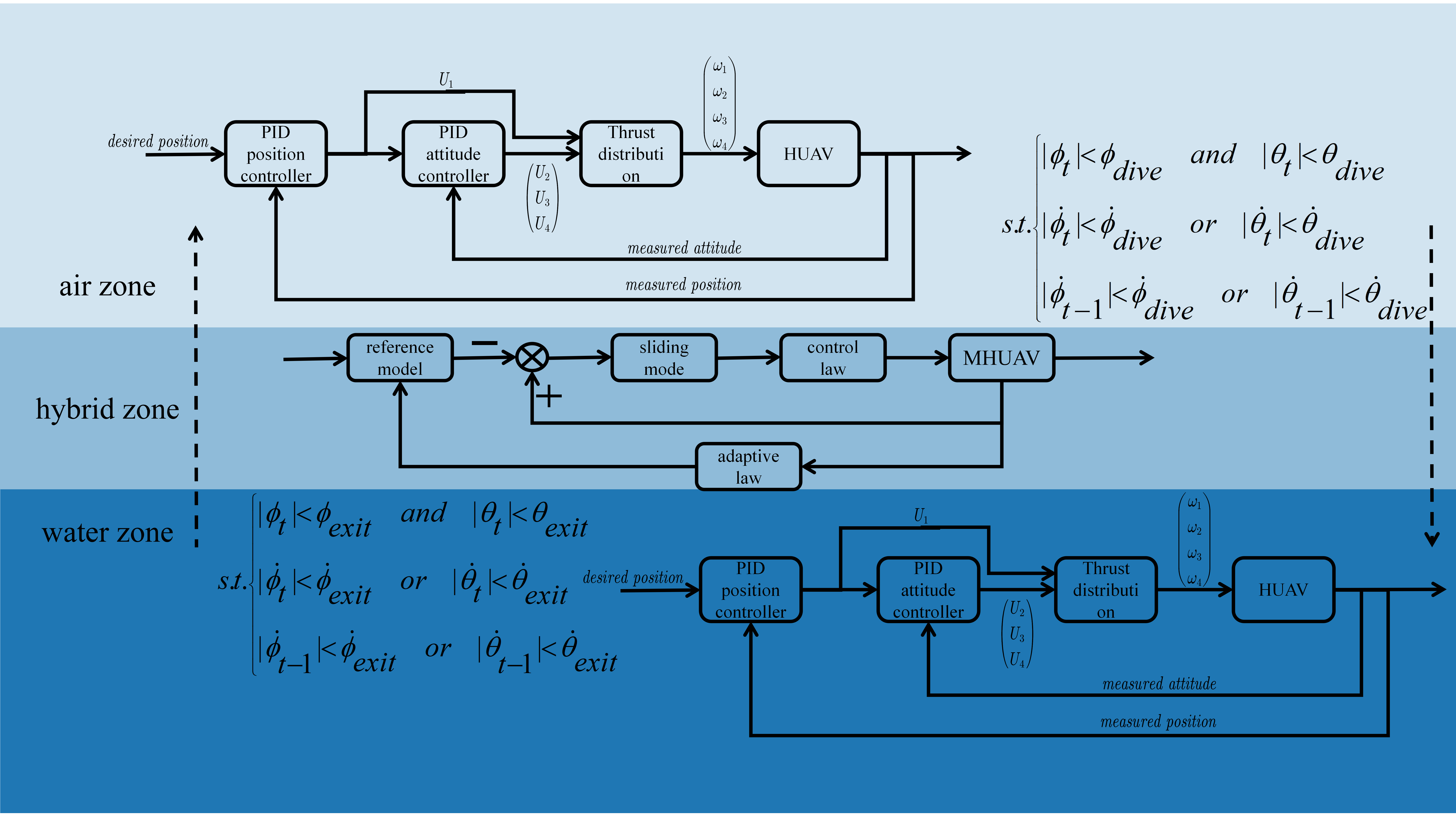}
    \caption{Control strategy}
    \label{control strategy}
\end{figure}

As shown in Fig. \ref{control strategy}, a three-tier control strategy is used, comprising an air control strategy (notated $S_A$), a water control strategy (notated $S_W$) and a hybrid control strategy (notated $S_H$), to accommodate the three dynamic models discussed in the previous section. Cascade PID control loops are utilized within $S_A$ and $S_W$ to achieve precise six-degree-of-freedom (6-DOF) control over position and attitude.\cite{chen2019system}
However, $S_H$ utilizes only the attitude control loop, as the primary focus in the hybrid zone is the longitudinal position and attitude stability of the vehicle, rather than controlling movement along the horizontal direction.
This hybrid structure aims to combine the precision and smoothness of PID control in predictable environments  with the robustness of TWSMC during the highly uncertain and disturbance-prone transition phase.
\subsection{Twisting sliding mode control(TWSMC)}

The twisting sliding mode control(TWSMC) method is employed within $S_H$ in order to achieve rapid response to errors, where the sliding mode surface is defined as follows:
\begin{equation}
    \mathbf{\sigma} = \dot{\mathbf{e}} + \mathbf{C} \circ |\mathbf{e}|^{\frac{1}{2}} \circ \mathrm{sgn}(\mathbf{e}) = \mathbf{0},
\end{equation}

where $\mathbf{e} = [e_z, e_\phi, e_\theta, e_\psi]^\top$ represents the position and angular error vector of the MHAUV, $\dot{\mathbf{e}}$ is its time derivative, and $\mathbf{C} = \mathrm{diag}(c_z, c_\phi, c_\theta, c_\psi)$ is a positive definite diagonal gain matrix. The operator $\circ$ denotes the Hadamard product (element-wise multiplication).
The derivative of the sliding mode surface is expressed as:
\begin{equation}\label{eq25}
    \dot{\mathbf{\sigma}} = \ddot{\mathbf{e}} + \frac{1}{2} \mathbf{C} \circ \dot{\mathbf{e}} \circ |\mathbf{e}|^{-\frac{1}{2}} = \ddot{\mathbf{q}} - \ddot{\mathbf{q}}_{\mathrm{ref}} + \frac{1}{2} \mathbf{C} \circ \dot{\mathbf{e}} \circ |\mathbf{e}|^{-\frac{1}{2}},
\end{equation}

where $\ddot{\mathbf{q}} = [\ddot{z}, \ddot{\phi}, \ddot{\theta}, \ddot{\psi}]^\top$ is the generalized acceleration vector and $\ddot{\mathbf{q}}_{\mathrm{ref}}$ is the reference acceleration vector.

Combining equations \ref{eq1}, \ref{eq1} and \ref{eq25}, the equivalent control terms are delivered as.
\begin{align}
\begin{split}
    T_{eq} &= \frac{m_v + m_a}{c_\theta s_\phi} \left[ \ddot{z_{ref}} - \frac{c_z \dot{e_z}}{2 |e_z|^{\frac{1}{2}}} + c_\theta \dot{\theta} u \right. \\
    &\quad + s_\theta \dot{u} - \dot{A_1} v - A_1 \dot{v} - \dot{A_2} w - A_2 Q u \\
    &\quad \left. + A_2 P v + A_2^2 \left( g_0 - \frac{B}{m_v} \right) + \frac{F_{dw}}{m_v + m_a} \right],
\end{split} \\
\begin{split}
    \tau_{x_{eq}} &= (I_{XX_v} + I_{XX_a}) \left[ -a_1 QR + A_2 W_G Q \right. \\
    &\quad - \dot{Q} A_3 - Q \dot{A_3} - \dot{R} A_4 - R \dot{A_4} \\
    &\quad \left. + \ddot{\phi_{ref}} - \frac{c_\phi \dot{e}_\phi}{2 |e_\phi|^{\frac{1}{2}}} \right] + M_{d\phi},
\end{split} \\
\begin{split}
    \tau_{y_{eq}} &= \frac{I_{YY} + I_{YY_a}}{c_\phi} [ \ddot{\theta_{ref}} - \frac{c_\theta \dot{e}_\theta}{2 |e_\theta|^{\frac{1}{2}}} - b_1 PR c_\phi\\
    &\quad - b_2 P W_G c_\phi + Q s_\phi \dot{\phi} + \dot{R} s_\phi + R c_\phi \dot{\phi} ] \\
    &\quad + M_{d\theta},
\end{split}\\
\begin{split}
    \tau_{z_{eq}} &= \frac{I_{ZZ}+I_{ZZ_a}}{A_6} [\ddot{\psi_{ref}} - \frac{c_\psi \dot{e_\psi}}{2 |e_\psi| ^\frac{1}{2}} - Q\dot{A_5}\\
    &- \dot{Q}A_5 - R\dot{A_6} - d_1PQA_6] + M_{d\psi},
\end{split}
\end{align}
\begin{equation}
\begin{gathered}
    A_1 = c_\theta s_\phi,\quad
    A_2 = c_\theta c_\phi,\quad
    A_3 = s_\phi t_\theta, \\
    A_4 = c_\phi t_\theta,\quad
    A_5 = \frac{s_\phi}{c_\theta},\quad
    A_6 = \frac{c_\phi}{c_\theta}, \\
    a_1 = \frac{I_{YY}-I_{ZZ}}{I_{XX}},\quad
    a_2 = \frac{I_{ZZM}}{I_{XX}}, \\
    b_1 = \frac{I_{ZZ}-I_{XX}}{I_{YY}},\quad
    b_2 = \frac{I_{ZZM}}{I_{YY}}, \\
    d_1 = \frac{I_{XX}-I_{YY}}{I_{ZZ}},
\end{gathered}
\end{equation}

where notations $A_{1\sim6}$, $a_{1\sim2}$, $b_{1\sim2}$ and $d_1$ are employed for simplification.

Applying the twisting control law, the convergence control term is defined as.
\begin{equation}\label{eq31}
\begin{gathered}
    u_{conv} = -r_1sgn(\sigma) - r_2sgn(\dot{\sigma}),\\
    r_1>r_2>0,
\end{gathered}
\end{equation}

where the switching magnitudes $r_1$ and $r_2$ are selected to satisfy the following conditions:
\begin{equation}
\begin{gathered}
    (r_1+r_2)K_m-C  > (r_1-r_2)K_M+C, \\
    (r_1-r_2)K_m > C,
\end{gathered}
\end{equation}

with constants $K_m$, $K_M$, and $C$ defined as constraints on the twice differentiable sliding variable $\sigma(t, x). $These constraints ensure that a second-order sliding mode appears in finite time, specifically achieving \(\sigma = \dot{\sigma} = 0.\)
Furthermore, these constants must satisfy:
\begin{equation}
\begin{gathered}
    0 < K_m < \frac{\partial}{\partial u}\ddot{\sigma} < K_M,\\
    \left|\ddot{\sigma}(u=0)\right|\leq C.
\end{gathered}
\end{equation}

In summary, the control law is derived by combining both equivalent and convergence control terms.

Smooth and stable switching between the control strategies ($S_A$, $S_H$, $S_W$) is critical. The switching logic is based on the vehicle's vertical position z relative to the hybrid zone boundaries (defined by $H_M=H/2$ and $H_m = -H/2$) and its vertical velocity $\dot{z}$. To prevent rapid oscillations near the boundaries (chattering), a small hysteresis or buffer zone $\delta z$ is introduced. Additionally, safe switching requires the vehicle's attitude and angular rates to be within acceptable limits before a switch occurs:
\begin{gather}
    |\phi|,|\theta| \leq a_M,\quad
    |\dot{\phi}|,|\dot{\theta}| \leq a_{vM},\\
    \begin{cases}
        if\quad z + \delta z=H_M \quad and \quad
        \begin{cases}
            if \quad \dot{z} \geq 0, \quad S_H\rightarrow S_A,\\
            if \quad \dot{z} < 0, \quad S_A\rightarrow S_H,
        \end{cases}\\
        if\quad z + \delta z=H_m \quad and \quad
        \begin{cases}
            if \quad \dot{z} > 0, \quad S_W\rightarrow S_H,\\
            if \quad \dot{z} \leq 0, \quad S_H\rightarrow S_W,
        \end{cases}
    \end{cases}
\end{gather}

where $a_M$ and $a_{vM}$ are predefined maximum allowable attitude angles and angular velocities. This ensures the vehicle is relatively stable before changing the control law. Ensuring global stability of such a switched system, especially with uncertain subsystems, can be complex and may require analysis using common Lyapunov functions or dwell-time conditions.

\subsection{Simulation and Comparison}
To evaluate the proposed hierarchical control strategy, simulations were conducted using Python. The simulation environment replicated the MHAUV dynamics described in Section 2, incorporating the hybrid model and the CFD-derived propeller characteristics. Environmental disturbances such ad waves and wind were initially excluded to isolate the controller performance during transition.

Three control strategies were compared: (1) Pure Cascade PID (using switched gains for air/water but PID throughout), (2) Pure TWSMC (using TWSMC throughout), and (3) the Proposed Hybrid TWSMC/PID controller. Performance was evaluated for tracking three different vertical trajectories designed to cross the air-water interface (z=0): a step change, a sinusoidal wave, and a cosine wave.

The physical parameters used in the simulation correspond to the prototype described in Section 4 and are listed in Table \ref{table simulation}.
\begin{table}
\caption{Parameters in the simulation}
    \centering
    \begin{tabular}{|c||c|}
    \hline
    \(m_v\) & 0.3 \(kg\) \\
    \hline
    \(m_a\) & 0.05 \(kg\) \\
    \hline
    \(A_{d0}\) & 0.02 \(m^2\) \\
    \hline
    \(V_0\) & 0.15 \(m^3\) \\
    \hline
    \(c_z\), \(c_\phi\), \(c_\theta\), \(c_\psi\) & 10 \\
    \hline
    \(r_1\) & 2500 \\
    \hline
    \(r_2\) & 1500 \\
    \hline
    \(K_P\) & 60 \\
    \hline
    \(K_I\) & 0.5 \\
    \hline
    \(K_D\) & 3000 \\
    \hline
    \(I_{XX}\), \(I_{YY}\) & 0.005 \(kgm^2\) \\
    \hline
    \(I_{ZZ}\) & 0.008 kg\(m^2\) \\
    \hline
    \end{tabular}
    
    \label{table simulation}
\end{table}

The simulation results indicate that our method demonstrates superior tracking performance across various media compared to PID, particularly when traversing downward through water. When the PID method descends through the water surface, the adjustment of the control variable is hindered by the need for accumulation in the integral component. This results in challenges when attempting to penetrate the water surface. In contrast, our proposed method is able to generate a more significant change in the control variable under identical error conditions, thereby facilitating a smoother transition through the water surface.

Furthermore, unlike TWSMC, our approach effectively mitigates buffeting in both air and aquatic environments by transitioning to a smoother PID controller within these domains.

\begin{figure} % Use appropriate placement specifiers like [!t], [!b], [!h]
    \centering % Center the group of subfigures
    % Use \subfloat for each subfigure.
    % Put the \label command *inside* the caption argument.
    % Use % at the end of lines to prevent unwanted spaces.
    \subfloat{%
        \includegraphics[width=0.5\linewidth]{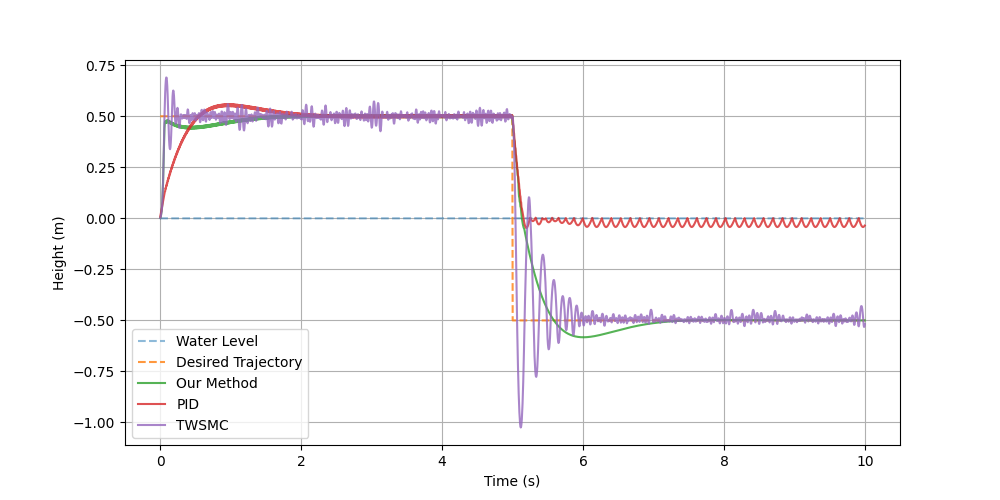}% Adjust width as needed (relative to \linewidth in figure*)
    }\hfill
    \subfloat{%
        \includegraphics[width=0.5\linewidth]{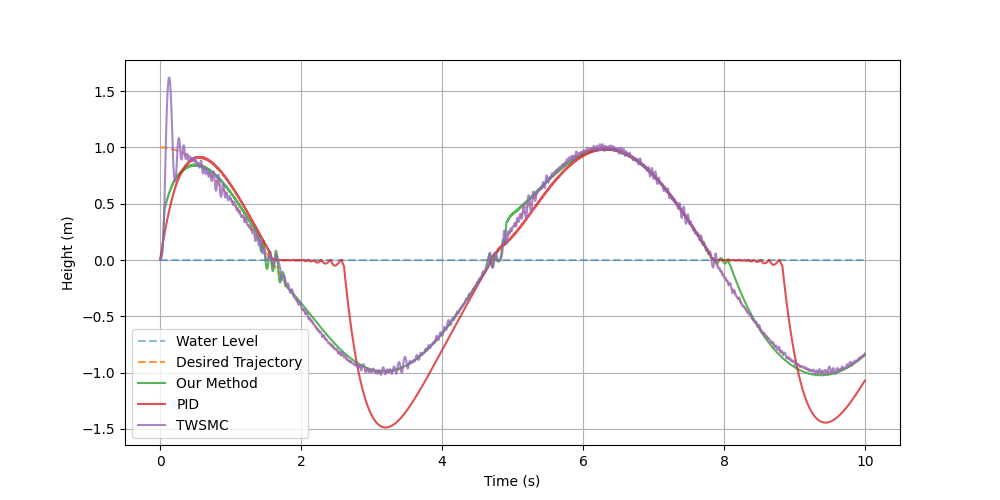}%
    }\hfill
    \subfloat{%
        \includegraphics[width=0.5\linewidth]{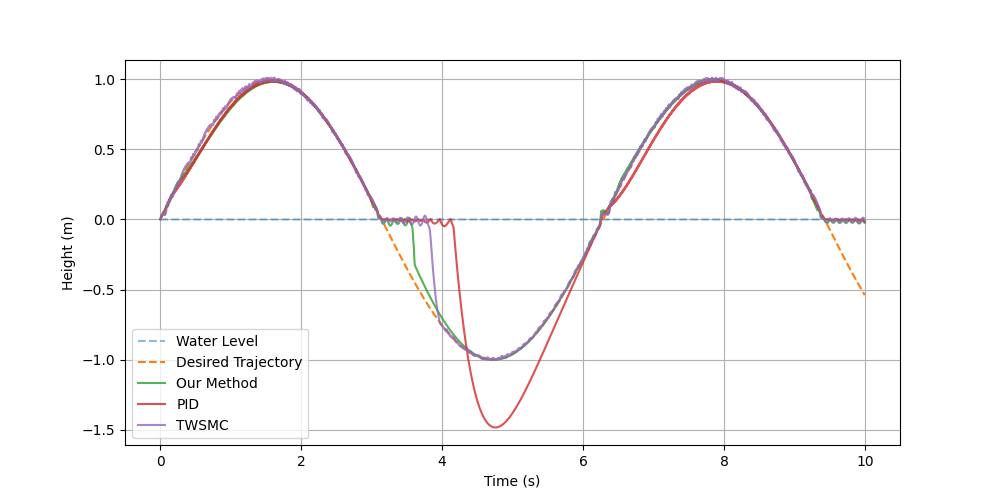}%
    }
    % Add an overall caption for the entire figure environment
    \caption{Simulation results for different desired curves.}
    % Add an overall label for the entire figure environment for cross-referencing
    \label{fig:simulation}
\end{figure}

\section{Prototype Design and Experiments}\label{sec:Prototype Design and Experiments}
To validate the proposed modeling and control framework in a real-world scenario, a low-cost MHAUV prototype was constructed and tested experimentally.\cite{bi2022nezha,liu2025wukong}

\begin{figure}
    \centering
    \includegraphics[width=0.8\linewidth]{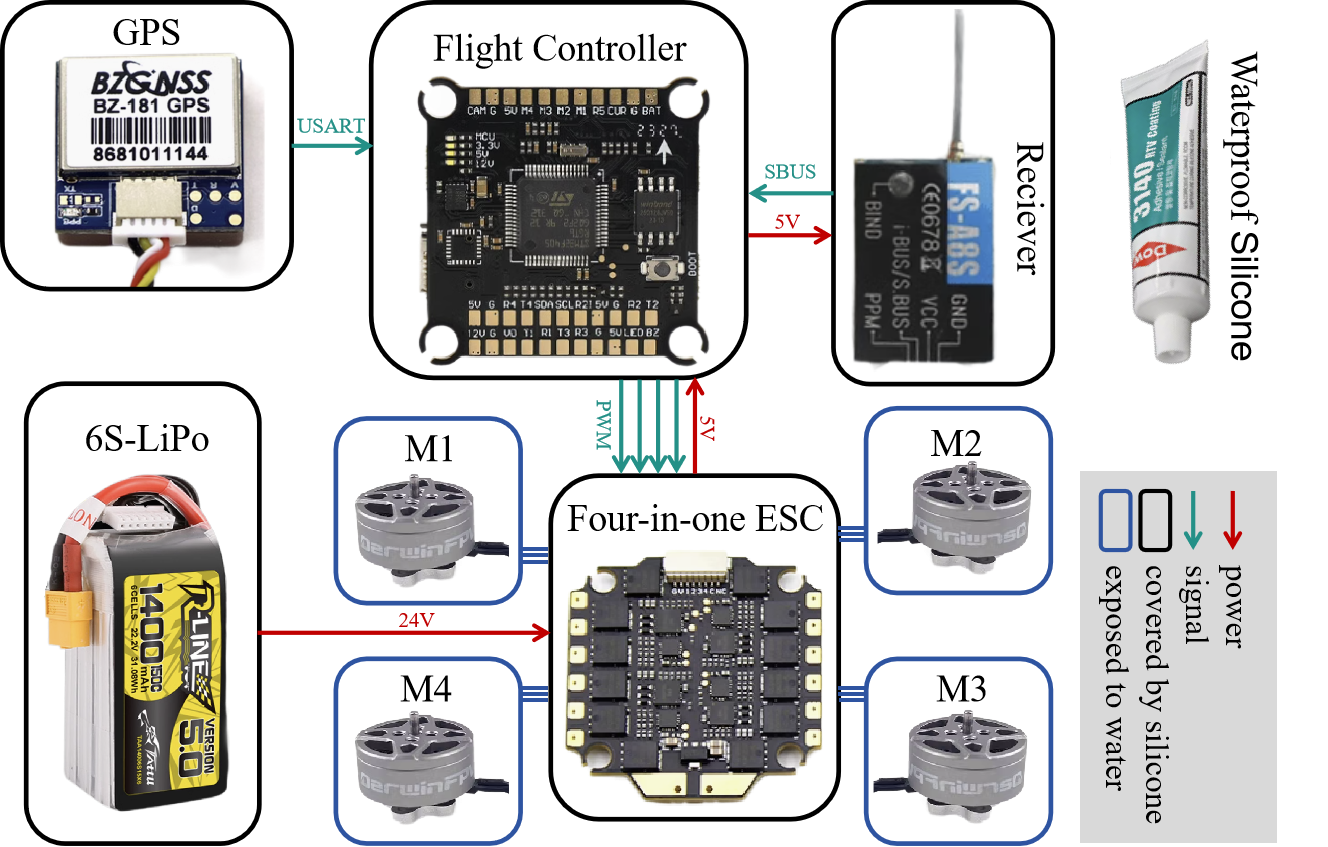}
    \caption{Structure of the low-cost prototype}
    \label{prototype design}
\end{figure}

For the purposes of cost-effectiveness and ease of reproducibility, the prototype was adapted from a first-person view (FPV) quadrotor aircraft. We developed our flight control program based on the Betaflight platform, utilizing the SPEDIX F405 flight control board. To prevent short circuits, we applied a layer of waterproof silicone to the surface of the control board.

\begin{figure}
    \centering
    \includegraphics[width=0.5\linewidth]{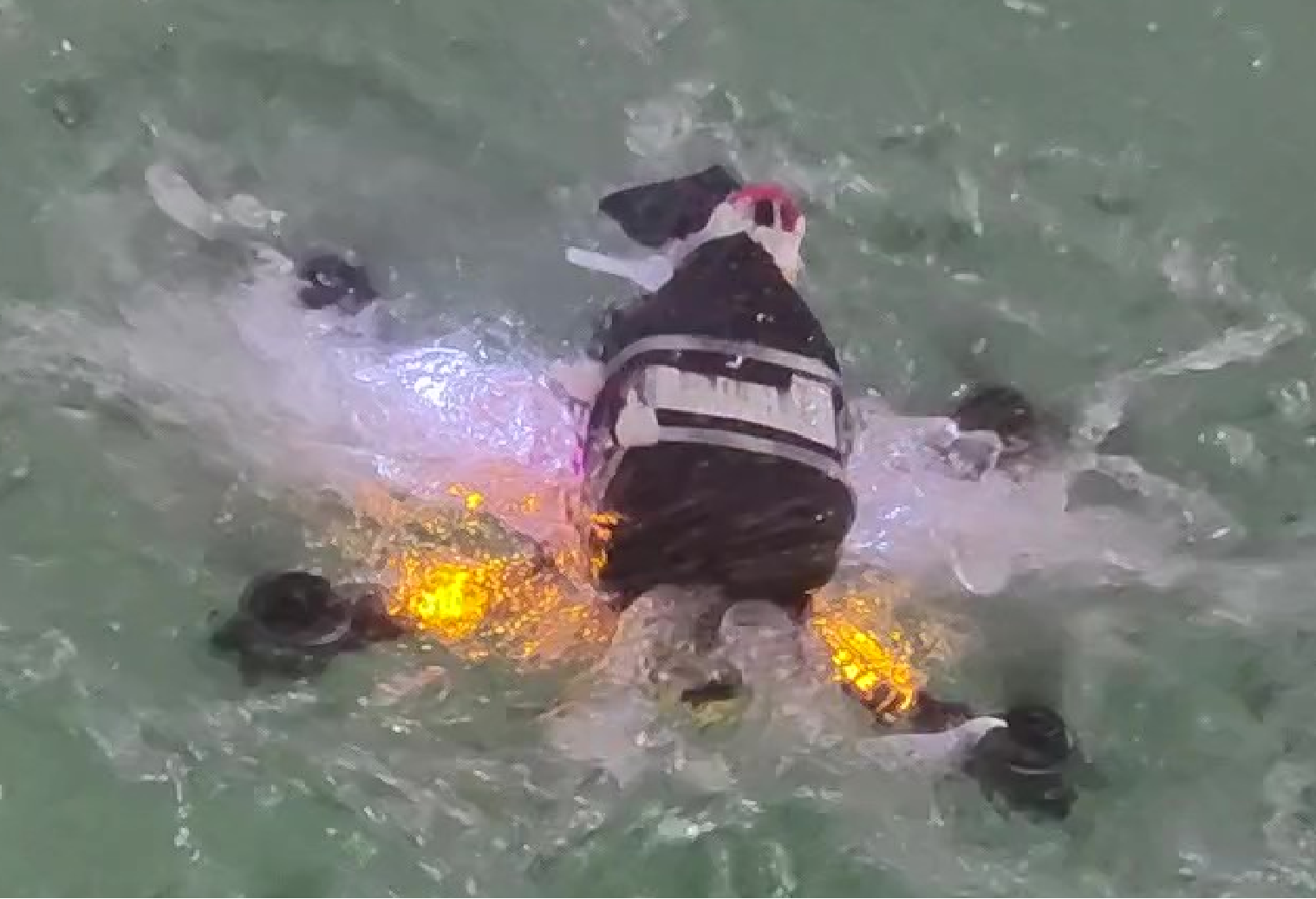}
    \caption{Experimental scene of prototype}
    \label{photo}
\end{figure}

The experiment was carried out under laboratory conditions. Initially, the prototype took off from a platform at the same height as the water surface($z=0$) and ascended to an altitude of 0.5$m$ above the water. Subsequently, it dived into the water and reached a depth of -0.5$m$. Finally, the prototype ascended, crossed back over the water surface, and returned to an elevation of 0.5 m.

\begin{figure}
    \centering
    \includegraphics[width=0.5\linewidth]{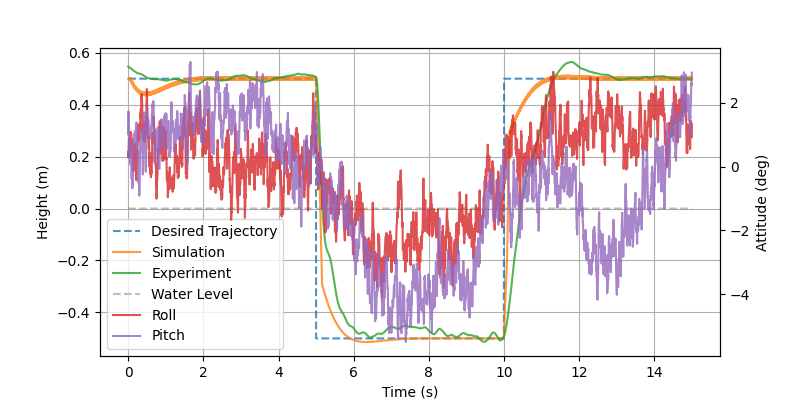}
    \caption{Comparison of experimental and simulation results}
    \label{ex result}
\end{figure}

\begin{figure} % Use appropriate placement like [!t], [!b], [!h]
    \centering % Center the image within the column
    % Include the first image, adjust width relative to column width
    \includegraphics[width=0.5\columnwidth]{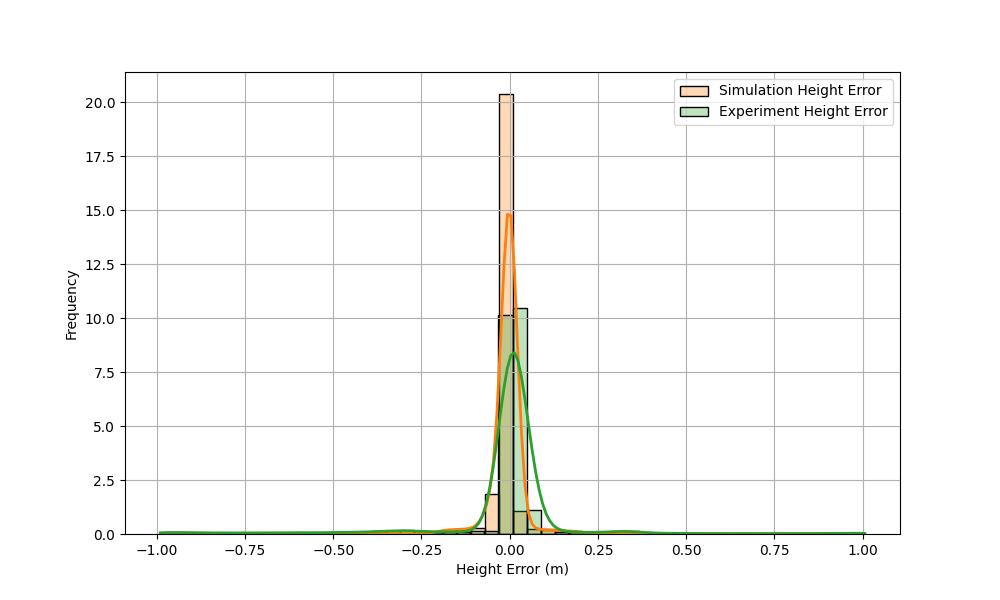}
    % Add caption and label for the first figure
    \caption{Height error distribution}
    \label{height error}
\end{figure}

\begin{figure} % Use appropriate placement like [!t], [!b], [!h]
    \centering % Center the image within the column
    % Include the second image, adjust width relative to column width
    \includegraphics[width=0.5\columnwidth]{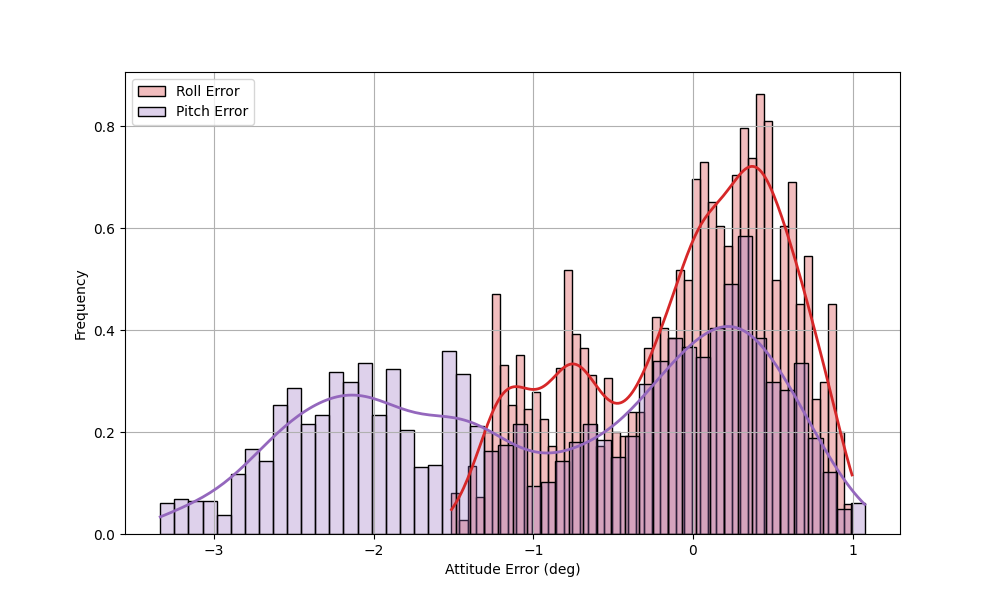}
    % Add caption and label for the second figure
    \caption{Altitude error distribution}
    \label{altitude error}
\end{figure}

The experimental results demonstrate the successful implementation of the modeling and control framework on the low-cost prototype.

Figure \ref{ex result} shows the measured vertical position (z) of the MHAUV during a representative water-crossing maneuver, compared to the desired trajectory. The vehicle successfully navigated the entire profile, including both water entry and water exit. The tracking performance appears qualitatively similar to the hybrid controller simulation results.

Quantitative analysis of the height error, shown in Figure \ref{height error}, confirms good tracking performance. In the steady-state portions (hovering in air or submerged), the height error remained below 0.1 meters. Transient errors occurred during the transitions, as expected due to the rapid dynamic changes, but the controller successfully stabilized the vehicle.

Figure \ref{altitude error} displays the roll and pitch angles recorded by the IMU during the experiment. The hierarchical controller, particularly the TWSMC active during transitions, effectively maintained attitude stability. Fluctuations were kept within a tight bound of approximately ±5 degrees throughout the entire maneuver, including the challenging entry and exit phases. This result is crucial, as it validates the simplifying assumption made in the hybrid dynamic model that the vehicle maintains a near-vertical orientation during transition.

The experiments successfully demonstrated the prototype's ability to perform repeated, seamless transitions between air and water. The combination of the hybrid dynamics model (incorporating CFD-based propeller data) and the hierarchical TWSMC/PID control strategy proved effective in managing the complex cross-domain dynamics on a low-cost hardware platform.

\section{Conclusion}\label{sec:Conclusion}
This paper has presented a systematic and experimentally validated framework for the modeling and control of a low-cost multirotor hybrid aerial-aquatic vehicle (MHAUV), directly addressing the significant challenges inherent in achieving seamless transitions between air and water environments.\cite{zeng2022review,yao2023review} The ability of HAUVs to operate across these distinct domains offers transformative potential for applications ranging from environmental monitoring and infrastructure inspection to search and rescue operations.\cite{xiang2015hybrid,bi2022nezha} Our work makes three principal contributions towards realizing robust and accessible cross-media robotic capabilities.

First, we developed a hybrid dynamics model to capture the complex physics governing the vehicle during the critical air-water transition phase. This model explicitly defines three operational zones: aerial, aquatic, and a transitional hybrid region, and incorporates key time-varying effects such as variable buoyancy, added mass, and depth-dependent fluid resistance.\cite{drews2014hybrid,mercado2018modeling}

The fidelity of this dynamic model was significantly enhanced by our second contribution: a systematic investigation of propeller thrust characteristics across the air-water interface using Computational Fluid Dynamics (CFD) simulations.\cite{alzubi2015evaluation,horn2019study,semenovnddevelopment} This analysis yielded a data-driven relationship characterizing the nonlinear variation of the thrust coefficient $C_T$ with propeller immersion depth $h$, providing crucial data for accurate force prediction during partial submersion—a factor often overlooked but vital for stable control.

Building upon this enhanced modeling foundation, we designed and implemented a hierarchical control architecture. This strategy strategically employs Twisting Sliding Mode Control (TWSMC) for robust attitude stabilization specifically during the highly dynamic and disturbance-prone medium transition phase.\cite{shtessel2017twisting,kochalummoottil2011adaptive,tahirovic2024twisting} TWSMC was chosen for its proven ability to provide strong robustness against uncertainties and external disturbances while mitigating the chattering associated with conventional SMC, making it well-suited for the transient dynamics of water surface penetration. This is complemented by cascade PID controllers, which ensure precise trajectory tracking and smooth operation within the more predictable homogeneous aerial and aquatic environments.\cite{chen2019system,zongcheng2022ga} This hybrid approach effectively balances the need for transient robustness with steady-state performance and control smoothness.

The integrated framework was experimentally validated using a prototype derived from a cost-effective commercial FPV quadrotor. The prototype successfully executed water-crossing maneuvers, achieving steady-state height tracking errors of less than 0.1 meters and maintaining attitude oscillations within 5 degrees. These results affirm the model's accuracy in predicting hydrodynamic-aerodynamic interactions and demonstrate the controller's capability to maintain stability amidst variations in fluid properties.

Future work will focus on extending the system's capabilities to handle more complex real-world conditions, primarily through the development of compensation strategies for wave disturbances by analyzing inertial-fluid coupling , and the creation of autonomous mission planning algorithms optimized for energy-efficient medium transitions to enable longer-duration operations and persistent autonomy.

% \begin{thebibliography}{1}
\bibliographystyle{IEEEtran}
\bibliography{IEEEabrv,citations}
% \end{thebibliography}

\newpage

\vfill

\end{document}